\documentclass[a4paper,11pt]{scrartcl}

\usepackage{graphicx} % Required for inserting images
\usepackage{amsmath,amsthm,amssymb} % Math
\usepackage[utf8]{inputenc} % Unicode input coding

\usepackage[margin=1.8cm,footskip=0.1cm]{geometry}

\usepackage[english]{babel} % Englsish

\usepackage[square,numbers]{natbib}
\usepackage[labelfont=bf, labelsep=colon, format=plain]{caption}

\usepackage{setspace}
\usepackage[section, below]{placeins}

\usepackage{lastpage}
\usepackage{fancyhdr}
\usepackage{url}

\usepackage{multicol}
\usepackage{multirow}

\newcommand{\myref}[2]{\textsc{#1}\ref{#2}}

\title{A GPU-enhanced workflow for non-Fourier SENSE
reconstruction}
\date{}
\author{}

\begin{document}

% Title Page 
\thispagestyle{empty}
\begin{center}
    \section*{A GPU-enhanced workflow for non-Fourier SENSE
reconstruction}
\end{center}
\noindent \textbf{Samuel Bianchi}\textsuperscript{1,*} (ORCID: 0009-0005-8255-9960)\\
\textbf{Klaas P. Pruessmann}\textsuperscript{1} (ORCID: 0000-0003-0009-8362) \\ \\
\noindent\textbf{1} Institute for Biomedical Engineering, ETH Zurich and University of Zurich, Switzerland\\ \\
\noindent\textbf{*} Corresponding author:\\ \\
\begin{tabular}{r l}
     \textbf{Name} & Dr. sc. Samuel Bianchi\\
     \textbf{Institute} & Institute for Biomedical Engineering  \\
     \textbf{Department} & Department of Information Technology and Electrical Engineering\\
     \textbf{University} & ETH Zürich \\ 
     \textbf{Address} & Gloriastrasse 35 \\
     & 8092 Zurich \\
     & Switzerland \\
     \textbf{E-mail} & bianchi@biomed.ee.ethz.ch
\end{tabular}
\newpage

\pagenumbering{arabic}

\section*{Abstract}
\textbf{Purpose:} Image reconstruction in challenging scenarios requires accurate characterizations of coil sensitivity profiles, local off-resonances ($B_0$) and effective encoding fields. Reconstruction methods utilizing all of this information rely on signal models that are not compatible with the classical Fourier/k-space interpretation of the coil data. Hence, the FFT and related techniques are no more applicable, rendering image reconstruction computationally demanding.\\ \\
\textbf{Methods:} This article contains a workflow for accurate sensitivity and $B_0$ mapping as well as other required processing steps. An implementation of non-Fourier SENSE reconstruction is provide that is well suited for execution on a GPU using the FFT. Important practical aspects like stopping criteria and sources of image artifacts are analyzed and documented.\\ \\
\textbf{Results:} Highly performant image reconstruction could be demonstrated on a 2D and 3D spiral dataset. These datasets contain trajectories featuring readout durations up to $71.5ms$ and undersampling factors up to $R=7$. Running the reconstruction on a GPU greatly boosts reconstruction speed. Stopping the reconstruction at the right moment is crucial for image quality. All methods included in this article are available in a public code repository.\\ \\
\textbf{Conclusion:} The provided implementation of non-Fourier SENSE reconstruction is highly performant. When it is executed on GPU, runtimes reach a duration feasible in practice. The presented workflow ensures robust and accurate computation of coil sensitive profiles and off-resonance maps.  \\ \\
\textbf{Keywords:} Reconstruction, SENSE, Higher-order fields, Non-FFT, GPU, Coil sensitivity,\\ $B_0$ map 
\newpage

\section*{Introduction}
Image reconstruction for magnetic resonance imaging (MRI) is still actively researched and developed. In challenging scenarios, an image reconstruction technique must be capable of: 
\begin{itemize}
    \item Achieving high parallel imaging performance. 
    \item Suppressing artifacts that originate from local off-resonances ($B_0$).
    \item Accounting for various encoding fields including arbitrary k-space trajectories and higher-order field components. 
\end{itemize}
Parallel imaging techniques \cite{Glockner2005,Deshane2012,Hamilton2017} reduce the amount of raw data that needs to be sampled by leveraging the inherent spatial encoding provided by an array of receiver coils. Popular choices include GRAPPA \cite{Griswold2002} and SENSE \cite{Pruessmann1999,Pruessmann2001} though many other techniques have been proposed \cite{Sodickson1997,Griswold2000,Blaimer2004}. Accurate characterization of the receiver coil array, e.g. via the GRAPPA kernel or sensitivity maps, is crucial for artifact-free image reconstruction. Furthermore, parallel imaging techniques inevitably reduce the signal-to-noise ratio (SNR) \cite{Pruessmann2006}. This reduction must be kept as minimal as possible. 
\\ When using long readouts, especially in combination with spiral trajectories \cite{Ahn1986,Bornert1999,Glover1998,Glover2012,Engel2018}, $B_0$ artifacts increasingly impair image fidelity \cite{Yudilevich1987,Schomberg1999,Block2005,Engel2018,Patzig2022}. Although several correction methods operating on already reconstructed images have been suggested \cite{Jezzard1995,Andersson2003}, full correction of these artifacts can only be achieved during image reconstruction if parallel imaging is used \cite{Patzig2022}. 
\\ Spirals, among many other non-Cartesian trajectories \cite{Irarrazabal1995,Zahneisen2012,Chauffert2014}, offer a high encoding efficiency \cite{Glover1998}. However, the reconstruction of non-Cartesian k-space data requires additional effort as the fast Fourier transform (FFT) cannot be applied directly \cite{Pruessmann2001,Jackson1991,Schomberg1995}. Additionally, due to the imperfections of the gradient system, the effective k-space trajectory may not traverse k-space on a Cartesian grid, even when intentionally designed to do so. Accounting for these deviations helps to ensure high image quality. Furthermore, higher-order components of the encoding fields \cite{Barmet2008} might become relevant in single-shot imaging \cite{Layton2013,Testud2015}, diffusion experiments \cite{Chan2014,Wilm2017,Ma2020}, when using high-performance gradient systems \cite{Weiger2018,Hennel2020}, or when implementing nonlinear spatial encoding schemes \cite{Tian2024}. NMR field cameras \cite{Barmet2008,Dietrich2016}, for example, can be used to measure dynamic field terms which are a representation of the effective encoding fields.
\\ Higher-order SENSE reconstruction \cite{Wilm2011} (see \myref{Figure }{fig_overveiw_plot} for an overview) can address all of the aforementioned issues by relying on a signal model that explicitly includes the following: Coil sensitivity maps ($S_\lambda(\mathbf{r})$), a $B_0$ map ($B_0(\mathbf{r})$), and dynamic field terms ($k_0(t),\dots,k_{P-1}(t)$). After discretization of the signal model, the image reconstruction is rendered as a large, linear inverse problem which is solved using the conjugate gradient (CG) method \cite{Hestenes1952}. Accordingly, higher-order SENSE reconstruction is a highly generic method, whose capabilities can be extended further simply by updating the signal model that is at its core. \\
It is important to note that higher-order SENSE reconstruction does not rely on the FFT, which is commonly used in MRI. Not depending on the FFT makes it such a generic and versatile algorithm because its implementation does not need to be compatible with using the FFT to map data from k-space to the image domain or vice versa. Accordingly, it does not rely on traditional Fourier-based interpretations of coil data. To reflect this property, rather than just its capacity for higher-order field terms, we propose the more descriptive term “non-Fourier SENSE reconstruction” which is used in the following.
\\ In practice, and in our experience, there are a few challenges associated with the use of non-Fourier SENSE reconstruction. Firstly, calculating accurate sensitivity and $B_0$ maps can be difficult. However, feeding inaccurate maps into the reconstruction will directly affect the achievable imaging performance. Secondly, non-Fourier SENSE reconstruction is computationally demanding. The discretization of the signal model yields a system matrix, referred to as encoding matrix ($\mathbf{E}$), which is too large to fit into memory \cite{Wilm2011}. As a consequence, all of its entries have to be recomputed in each iteration of the CG-method. This imposes a large computational overhead. Thirdly, the CG-method is an iterative solver for symmetric, linear systems of equations, and the number of iterations executed has a significant impact on the reconstructed image. Too few iterations result in image artifacts, while too many iterations result in noise amplification. Therefore, knowing when to stop iterating is of great practical relevance. Lastly, the final step of the reconstruction is to apply the k-space filter to the almost final image \cite{Pruessmann2001}. This filter suppresses image components whose reconstruction is ill-conditioned and, therefore, prone to strong noise amplification. The computation of this filter has to be done carefully to ensure maximal SNR of the final image.
\\ In this article, we present:
\begin{itemize}
    \item A workflow for the computation of sensitivity maps, $B_0$ maps and k-space filters that works robustly and ensures satisfactory reconstruction results. 
    \item A novel implementation of the reconstruction that is more memory efficient and faster specifically when running on a graphics processing unit (GPU). 
    \item  A discussion of how the number of CG iterations impacts image quality, as well as a comparison of stopping strategies. 
    \item Various examples to highlight the capabilities of the workflow and implementation. 
\end{itemize}
We hope that this work makes the non-Fourier SENSE reconstruction easily accessible to potential users. The accompanying git repository contains a fully functional and educational implementation of the proposed workflow. An example dataset is provided so that the code can be run without any need for adaptation or additional data collection.
\newpage

\section*{Methods and Material}\label{sec_mandm}
This section will guide through each of the processing steps included in the overview presented in \myref{Figure }{fig_overveiw_plot}, implementations of non-Fourier SENSE reconstruction and we assessed the impact of the number of CG iterations. Finally, we describe the dataset used for all analyses presented in this article. 

\subsection*{Computation of masks}
In this step, two masks are computed. The trusted mask ($M_T$) includes voxels with a sufficiently high SNR for a reliable estimation of the sensitivity maps. The reconstruction mask $M_R$ encompasses all voxels in which signal sources may be present. During non-Fourier SENSE reconstruction, non-zero voxel values are only permitted within $M_R$. This improves the conditioning of the inverse problem by reducing the number of unknown voxel values within the FOV. \\
After root sum of squares coil-combination of the prescan data, the magnitude image corresponding to the shortest echo time ($T_E$) iss bias-corrected for $B_1$ inhomogeneities. Using SPM’s (SPM12 r7771, \url{https://www.fil.ion.ucl.ac.uk/spm/}) unified segmentation \cite{Ashburner2005}, the bias-field and a probabilistic map of the “outer” tissue class are computed. $M_R$ is then obtained by excluding regions with a probability $<0.5$ of belonging to the “outer” class, keeping only the largest connected component, filling any remaining holes, and slightly dilating the result. $M_T$ is computed by thresholding the bias-corrected magnitude image. The threshold is determined as the location of the global minimum of the histogram of the logarithm of the bias-corrected magnitude image. 

\subsection*{Computation of sensitivity maps}
A sensitivity map $S_\lambda$ for each receiver-coil, indexed by $\lambda$, has to be computed within the area masked by $M_R$. Initial estimates $\hat{S}_\lambda$ are computed using singular value decomposition. For each voxel, the measured prescan data are arranged into a matrix with rows referring to receiver coils and columns to different echo times. The entries of the first left singular vector of this matrix are then used as the values of $\hat{S}_\lambda$ at voxel's location. Compared to other 
SVD-based methods for mapping receiver coil characteristics \cite{Brunner2016}, we make use of the temporal, and not spatial, dimension of the prescan data. \\
$\hat{S}_\lambda$ must be smoothed and extrapolated, particularly because the initial estimates are unreliable near the boundary of $M_R$ due to the absence of signal sources. We therefore propose to compute $S_\lambda$
such that it remains close to $\hat{S}_\lambda$ within $M_T$, while providing a smooth extrapolation of $\hat{S}_\lambda$ in regions included in $M_R$ but not in $M_T$. This objective can be achieved by modeling $S_\lambda$ as:
\begin{equation}\label{eq_sense_map_smoothing}
    \begin{pmatrix}
        \text{vec}(\hat{S}_\lambda) \\ 0 \\ 0 \\ 0
    \end{pmatrix}
    =
    \begin{pmatrix}
        \text{diag}(M_T) \\
        \sqrt{\alpha_S} \text{diag}(M_R) D_{2x} \\
        \sqrt{\alpha_S} \text{diag}(M_R) D_{2y} \\
        \sqrt{\alpha_S} \text{diag}(M_R) D_{2z}
    \end{pmatrix}
    \text{vec}(S_\lambda)
\end{equation}
The vectorization of a map is denoted by $\text{vec}()$, while $\text{diag}()$ constructs a matrix with the entries of a map on its diagonal. Applying $D_{2}$ to a vectorized map computes its second derivative along the x-, y-, or z-direction. $S_\lambda$ is the solution of this system of equations. The top line ensures consistency with the initial estimate and all subsequent lines ensure that $S_\lambda$ is smooth and extrapolated to cover the full $M_R$. $\alpha_S$ controls the strength of the smoothness penalty. \\
The above system of equations can be solved using the CG-method. We recommend implementing the system matrix using data structures optimized for sparse arrays. This facilitates the computation of a preconditioner, for instance using the incomplete Cholesky factorization, to lower the necessary number of CG iterations. Finally, the magnitude of $\hat{S}_\lambda$ and the magnitude-normalized $\hat{S}_\lambda$ are processed separately and subsequently recombined.

\subsection*{Computation of $B_0$ maps}
Complex-valued, coil-combined images are reconstructed from the prescan data using the sensitivity maps obtained in the previous step. The $B_0$ map is derived from the phase of these images. The phase of the image acquired at the lowest $T_E$ is set to $0$, and phase unwrapping is performed along the temporal (or echo) dimension. When accurate shimming has been performed, and the echo spacing is short enough, temporal phase unwrapping is sufficient for the purpose of $B_0$ mapping. Short echo spacing essentially enables temporal phase unwrapping in techniques like UMPIRE \cite{Robinson2014}.
However, if required or desired, spatial phase unwrapping \cite{Jenkinson2003,Schofield2003,AbdulRahman2007,Dymerska2021} may also be performed.\\
Following these steps, the initial estimate, $\hat{B}_0$, of the $B_0$ map is obtained by linearly fitting the slope of the phase evolution across the different echo times.
\begin{equation}
    \varphi_n(\mathbf{r}) =  \gamma  \hat{B}_0(\mathbf{r}) n \Delta T_E + \beta(\mathbf{r})\text{mod}(n,2) + \epsilon_n(\mathbf{r}) 
\end{equation}
Here, $n$ indexes images reffering to different echo times and $\Delta T_E$ denotes the echo spacing. The term $\beta(\mathbf{r})\text{mod}(n,2)$ accounts for small signal variations between even and odd echoes of unknown origin. $\epsilon_n(\mathbf{r})$ epresents the residual error of the linear fit. The corresponding standard error is given by $\varepsilon(\mathbf{r}) =\sqrt{ \frac{1}{N}  \sum_{n}{\epsilon^2_n(\mathbf{r})} }$ and used for smoothing the initially estimated map. \\
We chose a very similar linear model, as for the sensitivity maps, to smooth and extrapolate the $\hat{B_0}$.
\begin{equation}\label{eq_b0_map_smoothing}
    \begin{pmatrix}
        \text{vec}(\hat{B}_0) \\ 0 \\ 0 \\ 0
    \end{pmatrix}
    =
    \begin{pmatrix}
        I \\
        \sqrt{\alpha_B} \text{diag}(\varepsilon) D_{1x} \\
        \sqrt{\alpha_B} \text{diag}(\varepsilon) D_{1y} \\
        \sqrt{\alpha_B} \text{diag}(\varepsilon) D_{1z}
    \end{pmatrix}
    \text{vec}(B_0)
\end{equation}
Applying $D_1$ to a vectorized map yields its first derivative. By locally weighting the derivatives by $\varepsilon$ $B_0$ remains close to $\hat{B}_0$ wherever the initial estimation is accurate. Soothing predominately occurs where the initial estimation is inaccurate. Importantly this models ensures an edge-preserving processing of the $B_0$ map if the initial estimation is reliable. As different tissues show different magnetic susceptibilities, contributing to the local off-resonance, it has to be expected that the $B_0$ map has edges. 

\subsection*{Computation of k-space filters}
The k-space filter is applied to the reconstructed image by first transforming the image into k-space using FFT and subsequently multiplying it with the filter. The filtered image is then obtained by applying the inverse FFT, completing the reconstruction.
\\ According to their definition, the 1st-order dynamic field terms can be interpreted as k-space coordinates. The FFT yields a k-space representation of the image sampled on a Cartesian grid. Intuitively, the reconstruction of signal components corresponding to Cartesian grid points that lie far away from the k-space locations covered by the 1st-order dynamic field terms is ill-conditioned and prone to noise amplification. To mitigate this effect, a convex hull is fitted around all k-space coordinates derived from the 1st-order dynamic field terms, and Cartesian grid points lying inside this hull form the k-space filter. In practice, this can be achieved by computing the Delaunay triangulation of the k-space coordinates, and testing which Cartesian grid points lie inside a triangle or tetrahedron. 

\subsection*{Implementation of the higher-order SENSE reconstruction}
The original implementation of higher-order SENSE reconstruction \cite{Wilm2011} was based on the assumption that the encoding matrix ($\mathbf{E}$) cannot be stored in memory. Consequently, all elements of $\mathbf{E}$ were computed on the fly twice during each conjugate gradient iteration: Once for the multiplication of a vector by $\mathbf{E}$ and once for the multiplication by its conjugate transpose. While this approach is highly memory efficient, it is computationally expensive. \\
We present two alternative implementations that increase memory requirements but improve computational efficiency. Both approaches rely on two matrices that are smaller than $\mathbf{E}$ and enable operations equivalent to the multiplication by it. In addition, one implementation eliminates the need for any matrix recomputation, requiring sufficient memory but yielding faster runtimes. The second implementation recomputes a matrix once per CG iteration, thereby avoiding the need for double recomputation. The resulting reconstruction algorithms are depicted in \myref{Figure }{fig_recon_algorithms}. \\
Both algorithms need inputs: $\mathbf{\Sigma}$ containing the raw coil signal, $\mathbf{R}$ and $\mathbf{K}$ containing the spatial and temporal basis for phasing terms of the signal model, $\mathbf{S}$ containing the coil sensitivity maps, $\mathbf{j}$ containing the intensity correction, $\mathbf{f}$ containing the k-space filter and $N_{It}$ specifying the number of CG iterations. $\mathbf{S}$ and the matrix $\mathbf{P}$ are needed to define an operation equivalent to the multiplication by $\mathbf{E}$. $\mathbf{P}$ models phasing effects of the spatial encoding. 
\begin{equation}
    \mathbf{P}=e^{\circ i \mathbf{K}\mathbf{R}} \quad \mathbf{E}\mathbf{p} \equiv \mathbf{P}\left( \mathbf{S} \circ (\mathbf{p}\mathbf{1}_\Gamma^T)\right)
\end{equation}
Note that the operation on the right returns a matrix and not a vector as a result. Hence, both operations are equivalent up to reordering of the result. Similarly we have 
\begin{equation}
    \mathbf{E}^H \mathbf{\sigma} \equiv \left(\mathbf{S}^* \circ \left( \mathbf{P}^H \mathbf{\Sigma} \right) \right)\mathbf{1}_\Gamma^T
\end{equation}
However, because $\mathbf{P}$ is still a considerably large, dense and complex-valued matrix computing its conjugate transpose is time consuming. As $\mathbf{\Sigma}$ is of smaller size the operation equivalent to $(\mathbf{\sigma}^H\mathbf{E})^H$ is implemented on line $2$ of the higher-order SENSE reconstruction and on lines $2-10$ of the split higher-order SENSE reconstruction. This eliminates the computation of $\mathbf{P}^H$ fully. Operations equivalent to $\mathbf{E}^H\mathbf{E}\mathbf{\sigma}$ are implemented on line $7$ and on lines $16-24$ of the two algorithms.  \\
The key difference between the higher-order SENSE reconstruction and the split version is that for the first it is assumed that $\mathbf{P}$ fits into memory. It is computed once on line $1$, and never recomputed in the rest of the algorithm. In the split version blocks of $\mathbf{P}$ (referred to as $\mathbf{P}'$) are iteratively computed, applied and than overwritten in the next iteration. The index set $\mathcal{K}$ specifies the start indices of each block and is passed as an additional input to the split higher-order SENSE reconstruction. For both algorithms it is assumed that all other arrays fit into memory. \\
Clearly, both algorithms predominantly involve highly parallelization linear algebra operations. Accordingly, GPUs are expected to excel at running both algorithms. For a CUDA-based (Nvidia, Santa Clara CA, United States) implementation, it is recommended to implement the element-wise exponentiation with a dedicated CUDA kernel. Additionally, in the split higher-order SENSE reconstruction rows of $\mathbf{K}$ are frequently indexed. However, when arrays are stored in column-major order, performance can be enhanced by indexing columns of $\mathbf{K}^T$ and subsequently transposing the result (see lines $6$ and $20$). Therefore $\mathbf{K}$ is initially transposed on line $1$. The necessary adaptations for a row-major implementation are shown in \myref{Figure }{fig_recon_algorithms}.

\subsection*{Assessment of the impact of the number of CG-iterations}
The number of CG iterations has a substantial impact on image quality. An insufficient number of iterations leads to residual image artifacts, whereas an excessive number results in unnecessary noise amplification or the introduction of artifacts. To analyze these effects, we performed $250$ CG iterations. The evolution of the $\text{log()}$ of the residual norm versus the $\text{log()}$ of the solution norm (the L-curve \cite{Hansen1993}) and the structural similarity index measure (SSIM \cite{Zhou2004}) provide tools for assessing how the number of CG iterations influences image quality. It has been proposed to terminate the CG algorithm at the point of maximum curvature of the L-curve \cite{Gunawan2003}. Similarly, the peak SSIM over iterations might also provide a possible stopping criterion. 

\subsection*{Dataset description}
All data was acquired on a $3T$ Philips R2D2 system (Philips Healthcare, Best, The Netherlands) using the $16$-channel Skope NeroCam 3T (Skope Magnetic Resonance Technologies AG, Zurich, Switzerland) which features $16$ embedded 19F-NMR field probes for concurrent field monitoring. Prescan data was collected using a 2D spin-warp sequence \cite{Edelstein1980} (FOV: $219$x$219$x$44mm$, Resolution: $1$x$1$x$2mm$, Slice gap: $0mm$, $T_R =2.5s$, $T_E=4.6ms\cdots39.1ms$, $\Delta T_E=2.3ms$, $16$ echoes). To test the workflow one 2D and one 3D spiral-out dataset were acquires. The 2D spirals were acquire in single-shot manner (see \myref{Figure }{fig_spirals_2D} for details). Tilted hexagonal sampling (T-Hex \cite{Engel2021}) was implemented for 3D spirals (see \myref{Figure }{fig_spirals_3D} for details). 
\newpage

\section*{Results}
\subsection*{Image quality}
\myref{Figures }{fig_spirals_2D} and \myref{}{fig_spirals_3D} display the reconstructed images for the 2D spiral and 3D T-Hex datasets, respectively. Despite the use of demanding spiral trajectories featuring high undersampling factors and/or long acquisition durations ($T_{Aq}$), the results demonstrate consistently high image quality. No undersampling artifacts are visible. However, images computed based on the longest acquisitions (e.g., $R=2$ in \myref{Figure }{fig_spirals_2D}) exhibit moderate $B_0$ artifacts, visible as ringing or blurring near tissue-air interfaces. Higher-order field terms provided a marginal improvement of reconstruction performance, as shown in \myref{Figure }{fig_higher_order_examples}.

\subsection*{Runtime analysis}
The full runtime analysis of both suggested implementations of non-Fourier SENSE (see \myref{Figure }{fig_recon_algorithms}) is given in \myref{Table }{table_runtimes}. It is shown that the GPU consistently outperforms the CPU in terms of runtime. The initialization of $\mathbf{P}$ causes an initial computational overhead for the non-Fourier SENSE implementation. However, it outperforms the split implementation in all subsequent operations on both CPU and GPU. The first CG iteration is slower than the following iterations on the GPU due to additional initialization overheads. 

\subsection*{Impact of the number of CG iterations}
L-curves and the evolution of SSIM over CG iterations are depicted in \myref{Figure }{fig_cg_convergence}. All L-curves depict the expected L-shape. However, when assessing image quality visually, reconstruction would be stop many CG iterations before the point of maximum curvature of the L-curve is reached. \\
The SSIM-curves show that image quality quickly increases during the first $5$-$50$ iterations and decreases after an excessive amount of iterations. The example on the bottom of \myref{Figure }{fig_cg_convergence} shows that initially the image is dominated by artifacts. After a sufficient amount of iterations all artifacts have disappeared. When additional, unnecessary iterations are preformed, speckle noise is amplified and image quality decreases. The suggested stopping points lie relative close to the global maxima of the corresponding SSIM curve. 

\subsection*{Sensitivity and $B_0$ maps}
\myref{Figures }{fig_spirals_2D} and \myref{}{fig_spirals_3D} demonstrate that the computation and processing of sensitivity maps ($S_\lambda$) introduce no observable image artifacts. The effects of the smoothing and extrapolation algorithm are further detailed in \myref{Figure }{fig_sense_maps_examples}. The algorithm preserves the overall structure of the maps while effectively bridging noisy gaps and providing overall denoising. This leads to a slight increase in the SNR of the reconstructed images. \\
For the $B_0$ map, the smoothing and extrapolation algorithm perseveres edges and smooths the map selectively where it is noisy. This is shown in \myref{Figure }{fig_b0_maps_examples}. Not smoothing or masking noisy areas from the $B_0$ map results in artifacts close to the object's border. While applying a standard Gaussian filter produces a high-quality magnitude image, it distorts the phase image near the brain’s periphery. Additionally an artificial tissue contrasts is created in the phase image to compensate for the blurring of the $B_0$ map. \\
The effects of alternative reconstruction masks ($M_R$) are explored in \myref{Figure }{fig_maksing_examples}. Shrinking it to the trusted mask ($M_T$) introduces strong artifacts. By picking a too small $M_R$, the image can not properly account for all parts of the object which contain signal sources, resulting in the observed artifacts. Slightly dilating it has a moderate effect. However, extending it to cover the full FOV introduces new image artifacts as the inverse problem solved during the reconstruction becomes more ill-conditioned. 

\subsection*{K-space filter}
\myref{Figure }{fig_kspace_filter_examples} displays the full k-space content of a reconstructed image and explores the impact of varying the k-space filter's extent. Although the filter’s direct effect on the images is nearly imperceptible by eye, the difference images confirm that it effectively suppresses noise while leaving anatomical structures intact.
\newpage

\section*{Discussion}
Despite the sensitivity of spiral trajectories to local off-resonances and gradient system imperfections, high image quality was achieved using non-Fourier SENSE reconstruction. This was successfully demonstrated on a 2D and 3D dataset both featuring long readouts and high undersampling factors. For the first time, the impact the number CG iterations has on image quality has been assessed and documented. The proposed workflow ensures minimal image artifacts, truthful tissue contrast and high SNR. Furthermore, the suggested implementation offers an attractive reconstruction speed, particularly when run on a GPU. 

\subsection*{Implementation and runtime}
In contrast to the highly iterative original implementation \cite{Wilm2011}, both proposed implementations are optimized to utilize parallelizable linear algebra operations and eliminate the redundant recomputation of matrices $\mathbf{E}$ and $\mathbf{E}^H$ during each CG iteration. Note that, in many modern computation platforms linear algebra operations are parallelized in the background with minimal implementation overhead, and also on a CPU. The non-split variant removes matrix recomputation from the CG loop entirely. While a direct empirical comparison is not provided in this study, these improvements ensure that our implementations outperform the original one. \\
Running both algorithms on a GPU substantially reduced the runtime compared to execution on a CPU. Since the analysis was performed on a commercially available GPU, which is much cheaper than an MR system, it can be concluded that GPUs are a suitable and cost-effective platform for non-Fourier SENSE reconstruction. Although hardware specifications such as memory bandwidth, core count and available VRAM are important for overall performance, a thorough evaluation of their respective impact on is beyond the scope of this work. This is partly due to the black-box nature of the Matlab CUDA interface. However, performance could certainly be improved by implementing dedicated CUDA code. Nevertheless, GPUs with lots of memory are undoubtedly preferable, as they are more likely to enable the faster non-Fourier SENSE reconstruction not requiring any splitting of $\mathbf{P}$.

\subsection*{Number of CG iterations and stopping criteria}
Determining the optimal number of CG iterations is critical, as the stopping point impacts the achievable image quality. \myref{Figure }{fig_cg_convergence} provides the first assessment of the this effect in the context of iterative SENSE reconstruction. Clearly, an excessive amount of iterations diminishes image quality. The CG method returns an implicitly regularized solution of the system of equation. Specifically, the dimensionality of the Krylov subspace, which spans the basis of the solution, increases with each iteration \cite{Paquette2020}. Apparently, this popery of the CG method is crucial for ensuring high image quality. Otherwise, more iterations should lead to consistently increasing image quality. \\
Finding an automatic stopping criterion remains an open challenge for future research. Although the L-curve method was evaluated, it proved unsuitable for this specific application. In contrast, reference-based quality metrics like SSIM show greater promise. Notably, the adjustable parameters of SSIM could be calibrated such that reaching peak-SSIM could serve as a reliable stopping criterion for the CG method. However, finding the right SSIM calibration and proofing its generality, can only be achieved on a much larger dataset. 

\subsection*{Workflow}
The presented workflow for processing prescan data is intended as a suggestion rather than a strict requirement. Our objective is to demonstrate the workflow’s effectiveness, establish the level of processing quality needed for satisfactory reconstruction performance, and document modes of failure as well as how these result in visible artifacts. Notably, this workflow is also beneficial for FFT-based SENSE reconstruction \cite{Barmet2005}, which relies on the same masks, maps, and filter. \\
Masking was performed using standard bias correction and segmentation (provided by SPM12), followed by thresholding. Alternatives like N3 \cite{Sled1998} or N4ITK \cite{Tustison2010} are equally applicable for bias correction. While we utilized a simple histogram-based thresholding method with success, there are plenty of other methods to choose from \cite{Sezgin2004}. Notably, Otsu’s method \cite{Otsu1979} was not sensitive for this specific application in our tests. The reconstruction mask ($M_R$) could also be generated by dilating the trusted mask ($M_T$). As shown in \myref{Figure }{fig_maksing_examples}, a slightly oversized reconstruction mask does not significantly degrade performance. \\ \\
The smoothing and extrapolation of sensitivity maps successfully improved the SNR of the reconstructed images without introducing artifacts, as shown in \myref{Figure }{fig_sense_maps_examples}). This suggests that our signal model (\myref{Equation }{eq_sense_map_smoothing}) is a valid representation of the data. While sensitivity maps can also be estimated using a body-coil image, this approach requires an additional scan. Alternatively, the calibration step of ESPIRiT \cite{Uecker2014} could be used to calculate these maps if preferred. One challenge of the suggested smoothing method is that the resulting system of equations is poorly conditioned, which typically requires a high number of CG iterations to find an acceptable solution. However, implementing the suggested preconditioner effectively resolves this issue and improves convergence.  \\
Based on existing literature \cite{Yudilevich1987,Schomberg1999,Block2005,Engel2018,Patzig2022} and the analysis presented in \myref{Figure }{fig_b0_maps_examples} it is clear that accurate $B_0$ mapping is essential for high quality reconstruction. The comparison between the workflow and Gaussian-smoothed $B_0$ map shows that edges are an important property of $B_0$ maps and must be preserved during the smoothing. The penalization of the 1st derivative (see \myref{Equation }{eq_b0_map_smoothing}) successfully achieves edge preservation and the smoothing algorithm behaves similar to total variation denoising \cite{Rudin1992}(TVD). In fact, our model in \myref{Equation }{eq_b0_map_smoothing} can be derived by modifying the standard TV norm to penalize the squared $L_2$ norm of the image gradient and incorporating weighting based on the standard error. However, while typical TVD requires complex solvers \cite{Wahlberg}, our formulation results in a system of equations that can be solved efficiently using the CG method. Finally it has to be mentioned that other regularization methods for $B_0$ maps have been proposed \cite{Jenkinson2001}. However, they lack the required edge-preservation. \\
The k-space filter has received relatively little attention in SENSE reconstruction literature \cite{Pruessmann1999,Pruessmann2001,Barmet2005}. Although its impact on image quality is moderate (as shown in \myref{Figure }{fig_kspace_filter_examples}), it successfully improves the SNR, which is vital for noise-sensitive applications like functional MRI. As demonstrated in the provided code, the triangulation-based computation method is both simple and fast. Moreover, this approach is highly versatile and should be compatible with various other trajectories.

\subsection*{Beyond Fourier imaging}
Image reconstruction not relying on the FFT has several documented use-cases in literature \cite{Barmet2008,Wilm2011,Weiger2018,Hennel2020,Tian2024}. Apart from these use-cases, it has to be mentioned that maintaining the FFT at the core of SENSE reconstruction has become increasingly complex as capabilities of the method have been extended. Although gridding \cite{Jackson1991,Schomberg1995} enables the reconstruction of non-Cartesian k-space data \cite{Pruessmann2001}, it requires accurate sampling density estimation \cite{Pipe1999,Zwart2012} and kernel design \cite{Johnson2009}to perform optimally. Furthermore, achieving competitive runtimes with gridding necessitates specialized implementations\cite{Dale2001}. The integration of $B_0$ correction via multifrequency interpolation\cite{Man1997} introduces an additional free parameter: the number of frequency segments. When the correction of high-order field perturbations is required, approximating the encoding matrix via SVD becomes the only option to maintain FFT-compatibility \cite{Wilm2012}. Reconstruction performance is affected by the accuracy of the used approximation. Given that modern GPU technology facilitates rapid reconstruction without FFT, the reliance on the FFT’s computational efficiency may no longer justify the algorithmic complexity and potential sources of error it introduces. Non-Fourier SENSE offers a more direct, robust alternative that is less dependent on algorithmic design choices. \\ 
Furthermore non-Fourier SENSE reconstruction can easily be extended to address additional sources of artifacts. The matrix $\mathbf{P}$ accounts for global encoding effects included in the signal model. By incorporating prior knowledge of how magnetization behaves during the readout into this matrix, such as the $B_0$ map provided by a prescan, the reconstruction performance can be enhanced. As can be deduced from \myref{Figure }{fig_b0_maps_examples} the information included in $\mathbf{P}$ directly influences image contrast. While a deliberately blurred $B_0$ map creates artificial tissue contrast, the same mechanism could be used to accentuate desired tissue contrast. This shifts image contrast from something that is controlled by sequence parameters to something that can actively be enhanced during reconstruction. Early research has already successfully used this idea to enhance desired $T_2^*$-contrast while suppressing artifacts originating from dropout regions \cite{Bianchi2025}.
\newpage

\section*{Conclusions}
Non-Fourier SENSE is a powerful reconstruction technique, that does not rely on the FFT, enabling high-quality image reconstruction even in highly challenging scenarios. By leveraging GPU acceleration, the computational runtime is significantly reduced, making non-FFT-based methods practical for real-world applications. To ensure optimal image quality, it is essential to terminate the CG iterations at the right moment. Finally, the proposed workflow for prescan data works robustly and satisfactory results are consistently ensured. 

\section*{Data Availability Statement}
A Matlab implementation of the workflow and reconstruction algorithms, including a dataset, can be found under: \url{https://gitlab.ethz.ch/sbianchi/example_code_nonfourier_sense}. The code is published under the GNU AFFERO General Public License v3.0 (AGLPv3). The use of the dataset is regulated under the accompanying data usage agreement.

\section*{Acknowledgments}
We thank all volunteers who participated in the experiments conducted for this study.

\section*{Financial disclosure}
None reported.

\section*{Conflict of interest}
The authors declare no potential conflict of interests.

\newpage

\bibliography{references}
\newpage

\section*{Figures and tables}

\begin{figure*}
\centerline{\includegraphics[width=17cm,height=18.9cm]{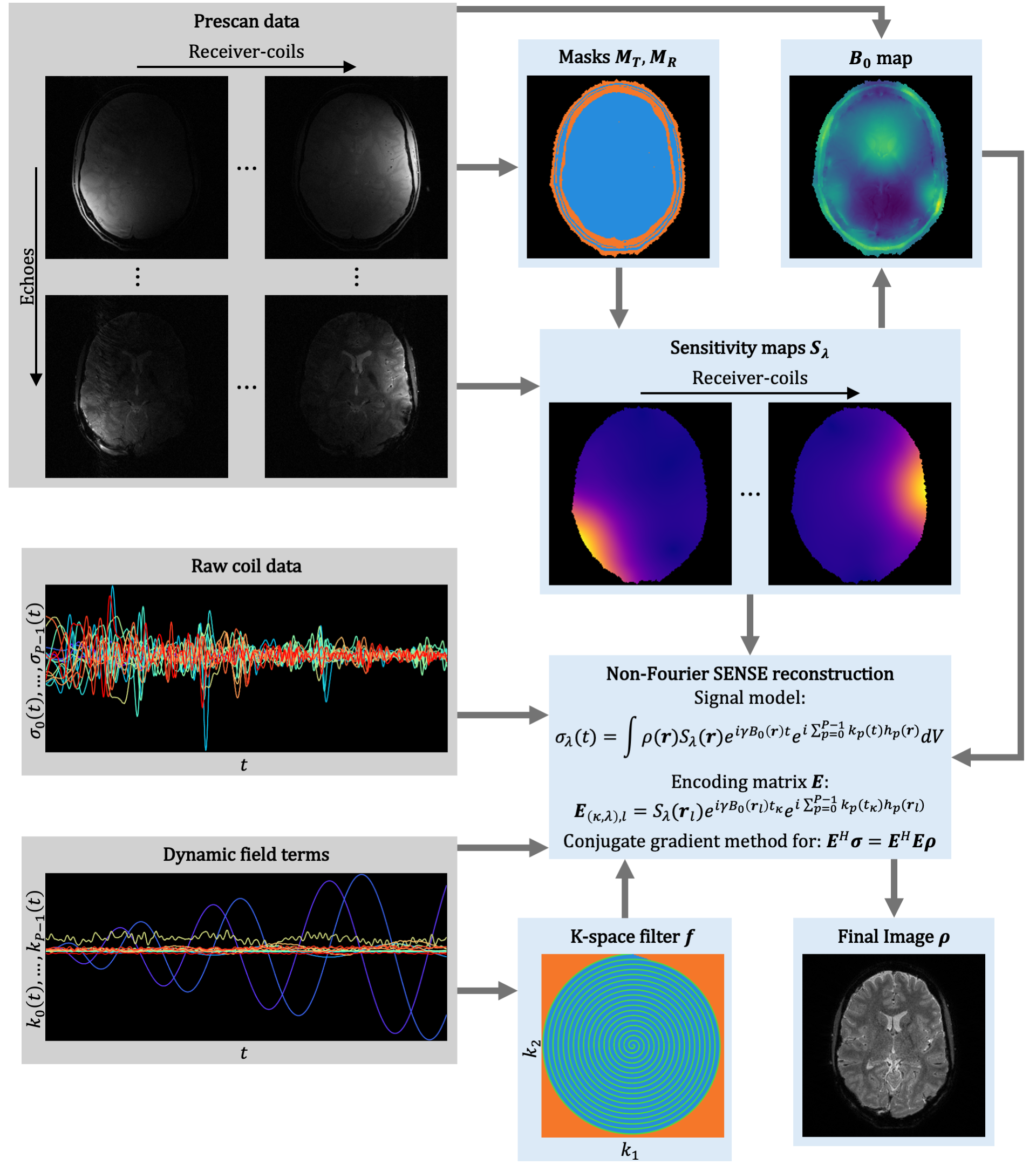}}
\caption{Overview of all the necessary data and processing steps for non-Fourier SENSE reconstruction. Prescan data are used to compute a trusted mask ($M_T$, blue) and a reconstruction mask ($M_R$, red). The trusted mask contains all voxels with a sufficiently high signal-to-noise ratio (SNR) to compute sensitivity maps $S_\lambda$, and the reconstruction mask contains all voxels where signal sources (spins) may have been present. Both masks and the prescan data are then used to compute sensitivity maps. After coil combination, using the sensitivity maps, a $B_0$ map can also be computed. Dynamic field terms and raw coil data complement all terms which need to be known for image reconstruction according to the signal model. In non-Fourier SENSE reconstruction, the signal model is discretized, yielding a linear system of equations that is solved using the conjugate gradient method. After reconstruction the image is filtered using the k-space filter $\mathbf{f}$. The elements in this overview which are addressed in this article are highlighted in light blue. Specifically addressed are the masking, the computation sensitivity and $B_0$ maps, the computation of the k-space filter, the implementation of the reconstruction itself and  the assessment of the reconstructed images ($\rho$). \label{fig_overveiw_plot}}
\end{figure*}
\newpage

\begin{figure*}
\scalebox{0.56}{
\begin{tabular}{ll}
% full sense 
\begin{tabular}{l r l l}
    \multicolumn{4}{l}{\textbf{Higher-order SENSE reconstruction}}  \\ \hline \hline  
      & Input: & $\mathbf{\Sigma}$, $\mathbf{R}$, $\mathbf{K}$,  $\mathbf{S}$, $\mathbf{j}$, $\mathbf{f}$, $N_{It}$ & \\ 
      & Output: & $\mathbf{\rho}$ &  \\ \hline \\[-2.5ex] 
      & 1: & $\mathbf{S} = \mathbf{S} \circ (\mathbf{j}\mathbf{1}_\Gamma^T)$ & Intensity correction \\
      & 2: & $\mathbf{P} = e^{\circ i \mathbf{K}  \mathbf{R}}$ & Initialized $\mathbf{P}$ \\
      & 3: & $\mathbf{p} = \left( \mathbf{1}_\Gamma^T \left(\left( \mathbf{\Sigma}^H \mathbf{P} \right) \circ \mathbf{S}^T \right) \right)^H$  & Compute $\mathbf{E}^H \mathbf{\sigma}$ \\
      & 4: & $\mathbf{r} = \mathbf{p}$ & Initialize residual \\
      & 5: & $\mathbf{\rho} = \mathbf{0}_{L}$ & Initialize solution \\
      & 6: & $n = 1$ & Iteration counter \\
      & 7: & while $n \leq N_{It}$ do & Loop over CG iterations \\ 
      & 8: & $\quad$ $\mathbf{q} = \left( \mathbf{1}_\Gamma^T\left(\left(\left( \mathbf{P}\left(\mathbf{S} \circ \left(\mathbf{p}\mathbf{1}_\Gamma^T\right)\right)\right)^H \mathbf{P} \right) \circ \mathbf{S}^T \right)\right)^H$ & Compute $\mathbf{E}^H \mathbf{E} \mathbf{p}$ \\
      & 9: & $\quad$ $\alpha = \mathbf{r}^H \mathbf{r}$ & Compute update rates \\
      & 10: & $\quad$ $\beta = \mathbf{p}^H \mathbf{q}$ & \\
      & 11: & $\quad$ $\mathbf{\rho} = \mathbf{\rho} + \frac{\alpha}{\beta} \mathbf{p}$ & Update solution \\
      & 12: & $\quad$ $\mathbf{r} = \mathbf{r}- \frac{\alpha}{\beta} \mathbf{q}$ & Update residual \\
      & 13: & $\quad$ $\beta =\alpha$ & Prepare next iteration \\
      & 14: & $\quad$ $\alpha = \mathbf{r}^H \mathbf{r}$ & \\
      & 15: & $\quad$ $\mathbf{p} =\mathbf{r}+ \frac{\alpha}{\beta} \mathbf{p}$ & \\
      & 16: & $\quad$ $n = n + 1$ & \\
      & 17: & end while & Stop CG itearations\\
      & 18: & $\mathbf{\rho} = \mathbf{\rho} \circ \mathbf{j}$ & \\
      & 19: & $\mathbf{\rho} = \text{IFFT}\left( \text{FFT}(\mathbf{\rho}) \circ \mathbf{f} \right)$ & Apply k-space filter \\ \hline
\end{tabular}
& 
\multirow[b]{-11.94}{*}{
\begin{tabular}{l r l l}
    \multicolumn{4}{l}{\textbf{Column-major, split higher-order SENSE reconstruction}}  \\ \hline \hline 
      & Input: & $\mathbf{\Sigma}$, $\mathbf{R}$, $\mathbf{K}$,  $\mathbf{S}$, $\mathbf{j}$, $\mathbf{f}$, $N_{It}$,  & \\ 
      & &  $\mathcal{K}=\{\kappa_1,\kappa_2,\dots \}$ & \\ 
      & Output: & $\mathbf{\rho}$ & \\ \hline  \\[-2.5ex] 
      & 1: & $\mathbf{S} = \mathbf{S} \circ (\mathbf{j}\mathbf{1}_\Gamma^T)$ & Intensity correction \\
      & 2: & $\mathbf{K} = \mathbf{K}^T$ & Ensure fast indexing \\
      & 3: & $\mathbf{\Sigma} = \mathbf{\Sigma}^{H}$  & Prepare $\mathbf{\Sigma}$ for $\mathbf{E}^H\mathbf{\sigma}$ \\
      & 4: & $\mathbf{p} = \mathbf{0}_{\Gamma,L}$ & Prepare storage for result\\
      & 5: & $m=1$ & \\
      & 6: & while $m \leq |\mathcal{K}|-1$ do & Iterative computation of $\mathbf{E}^H\mathbf{\sigma}$\\
      & 7: & $\quad$ $\mathbf{P}' = e^{\circ i \mathbf{K}^T_{(:,\kappa_{m}:\kappa_{m+1}-1)} \mathbf{R}}$ & Compute part of the full $\mathbf{P}$ \\
      & 8: & $\quad$ $\mathbf{p} = \mathbf{p} + \mathbf{\Sigma}_{(:,\kappa_{m}:\kappa_{m+1}-1)} \mathbf{P}'$ & Apply it and sum result \\
      & 9: & $\quad$ $m = m+1$ & \\
      & 10: & end while & \\
      & 11: & $\mathbf{p} = \left(\mathbf{1}_\Gamma^T\left(\mathbf{p}\circ\mathbf{S}^T\right)\right)^H$ & Reverse sensitivity weighting \\
      & 12: & $\mathbf{r} = \mathbf{p}$ & Initialize residual \\
      & 13: & $\mathbf{\rho} = \mathbf{0}_L$ & Initialize solution \\
      & 14: & $n=1$ & \\
      & 15: & while $n\leq N_{It}$ do & Loop over CG iterations \\
      & 16: & & Lines 16-24 implement $\mathbf{E}^H \mathbf{E} \mathbf{\sigma} $ \\ 
      & 17: & $\quad$  $\mathbf{Q} = \mathbf{S} \circ \left( \mathbf{p} \mathbf{1}_\Gamma^T \right)$& Forward sensitivity weighting \\
      & 18: & $\quad$ $\mathbf{Q}'=\mathbf{0}_{\Lambda,L}$ & Buffer for iterative multiplication \\
      & 19: & $\quad$ $m = 1$ & \\
      & 20: & $\quad$ while $m \leq |\mathcal{K}|-1$ do & Iterative computation of $\mathbf{P}^H \mathbf{P} \mathbf{Q}$\\
      & 21: & $\quad$ $\quad$ $\mathbf{P}' = e^{\circ i \mathbf{K}^T_{(:,\kappa_{m}:\kappa_{m+1}-1)} \mathbf{R}}$ & Compute part of the full $\mathbf{P}$ \\
      & 22: & $\quad$ $\quad$ $\mathbf{Q}' = \mathbf{Q}' + (\mathbf{P}' \mathbf{Q})^H \mathbf{P}'$ & Apply it and sum result \\
      & 23: & $\quad$ $\quad$ $m = m+1$ & \\
      & 24: & $\quad$ end while & \\
      & 25: & $\quad$ $\mathbf{Q} = \left(\mathbf{1}_\Gamma^T\left(\mathbf{Q}'\circ\mathbf{S}^T\right)\right)^H$ & Reverse sensitivity weighting\\
      & 26: & $\quad$ $\alpha = \mathbf{r}^H \mathbf{r}$ & Compute update rates \\
      & 27: & $\quad$ $\beta = \mathbf{p}^H \mathbf{Q}$ & \\
      & 28: & $\quad$ $\mathbf{\rho} = \mathbf{\rho} + \frac{\alpha}{\beta} \mathbf{p}$ & Update solution \\
      & 29: & $\quad$ $\mathbf{r} = \mathbf{r}- \frac{\alpha}{\beta} \mathbf{Q}$ & Update residual \\
      & 30: & $\quad$ $\beta =\alpha$ & Prepare next iteration \\
      & 31: & $\quad$ $\alpha = \mathbf{r}^H \mathbf{r}$ & \\
      & 32: & $\quad$ $\mathbf{p} =\mathbf{r}+ \frac{\alpha}{\beta} \mathbf{p}$ & \\
      & 33: & $\quad$ $n = n + 1$ & \\
      & 34: & end while & Stop CG itearations\\
      & 35: & $\mathbf{\rho} = \mathbf{\rho} \circ \mathbf{j}$ & \\
      & 36: & $\mathbf{\rho} = \text{IFFT}\left( \text{FFT}(\mathbf{\rho}) \circ \mathbf{f} \right)$ & Apply k-space filter \\ \hline
      \\
      \multicolumn{4}{l}{\textbf{Row-major, split higher-order SENSE reconstruction (adaptions only)}}  \\ \hline \hline 
      & & & Inputs/Output does not change \\ \hline
      & 1: & & Transposition of $\mathbf{K}$ not needed \\
      & 6: & $\quad$ $\mathbf{P}' = e^{\circ i \mathbf{K}_{(\kappa_{m}:\kappa_{m+1}-1,:)} \mathbf{R}}$ & Change of indexing, no transposition of $\mathbf{K}$ \\
      & 20: & $\quad$ $\quad$ $\mathbf{P}' = e^{\circ i \mathbf{K}_{(\kappa_{m}:\kappa_{m+1}-1,:)} \mathbf{R}}$ &  Change of indexing, no transposition of $\mathbf{K}$ \\ \hline
\end{tabular}
}
\\ &  \\ 
\begin{tabular}{l r l l}
    \multicolumn{4}{l}{\textbf{Description of variables and operations}}  \\ \hline \hline 
    & Inputs: & $\mathbf{\Sigma}_{(\kappa,\lambda)} = \sigma_\lambda(t_\kappa)$ & $K\times\Gamma$ matrix with raw coil signal \\ 
    & & $\mathbf{R}_{(1,l)} = B_0(\mathbf{r}_l)$ & $P+1 \times L$ matrix with spatial basis \\
    & & $\mathbf{R}_{(p+2,l)} = h_p(\mathbf{r}_l)$ & for phasing terms \\ 
    & & $\mathbf{K}_{(\kappa,1)} = t_\kappa$ & $K\times P + 1$ matrix with temporal basis\\
    & & $\mathbf{K}_{(\kappa,p+2)} = k_p(t_\kappa)$ & for phasing terms \\ 
    & & $\mathbf{S}_{(l,\lambda)} = S_\lambda(\mathbf{r}_l)$ & $L \times \Gamma$ matrix with sensitivity maps\\ 
    & & $\mathbf{j}$ & Intensity correction as a vector \\
    & & $\mathbf{f}$ & K-space filter as vector  \\
    & & $N_{It}$ & Number of CG iterations \\
    & & $\mathcal{K}$ & Set of start-indices ($\kappa$) of blocks of $\mathbf{K}$ \\ \hline
    & Output: & $\mathbf{\rho}$ & Reconstructed image \\ \hline
    & Arrays: & $\mathbf{0}_{L}$  & $L \times 1$ column vector filled with $0$s \\ 
    &       & $\mathbf{0}_{\Gamma,L}$  & $L \times \Gamma$ matrix filled with $0$s \\ 
    & & $\mathbf{1}_\Gamma$ & $\Gamma \times 1$ column vector filled with $1$s \\ \hline
    & Dimensions: & $L$ & Number of voxels \\
    & & $K$ & Number of k-space samples \\
    & & $\Gamma$ & Number of receiver coils \\
    & & $P$ & Number of dynamic field terms \\ \hline
    & Operations: & $\mathbf{A}^T$ & Transpose of $\mathbf{A}$ \\ 
    & & $\mathbf{A}^H$ & Conjugate transpose of $\mathbf{A}$ \\
    & & $\mathbf{A} \circ \mathbf{B}$ & Element-wise multiplication of $\mathbf{A}$ and $\mathbf{B}$ \\
    & & $e^{\circ \mathbf{A}}$ & Element-wise exponentiation of $\mathbf{A}$ \\
    & & $\mathbf{A}_{(m,n)}$ & Indexing of one element of $\mathbf{A}$ \\
    & & $\mathbf{A}_{(:,n_1:n_2)}$ & Indexing of all rows and columns $n_1$ to $n_2$ \\ 
    & & $\text{FFT()}$ & Fast Fourier transformation \\
    & & $\text{IFFT()}$ & Inverse fast Fourier transformation \\ \hline
\end{tabular}
& 
\end{tabular}}

\caption{Top-left: Non-Fourier SENSE reconstruction if enough memory for all arrays is available. Top-right: Column-major non-Fourier SENSE reconstruction if not enough memory for the matrix $\mathbf{P}$ is available. All other arrays are expected to fit into memory. Bottom-left: Descriptive information for both algorithms. Bottom-right: Necessary adaptions of the column-major non-Fourier SENSE reconstruction for an implementation suited for row-major memory layout.  \label{fig_recon_algorithms}}
\end{figure*}
\newpage

\begin{figure*}
\centerline{\includegraphics[width=17cm]{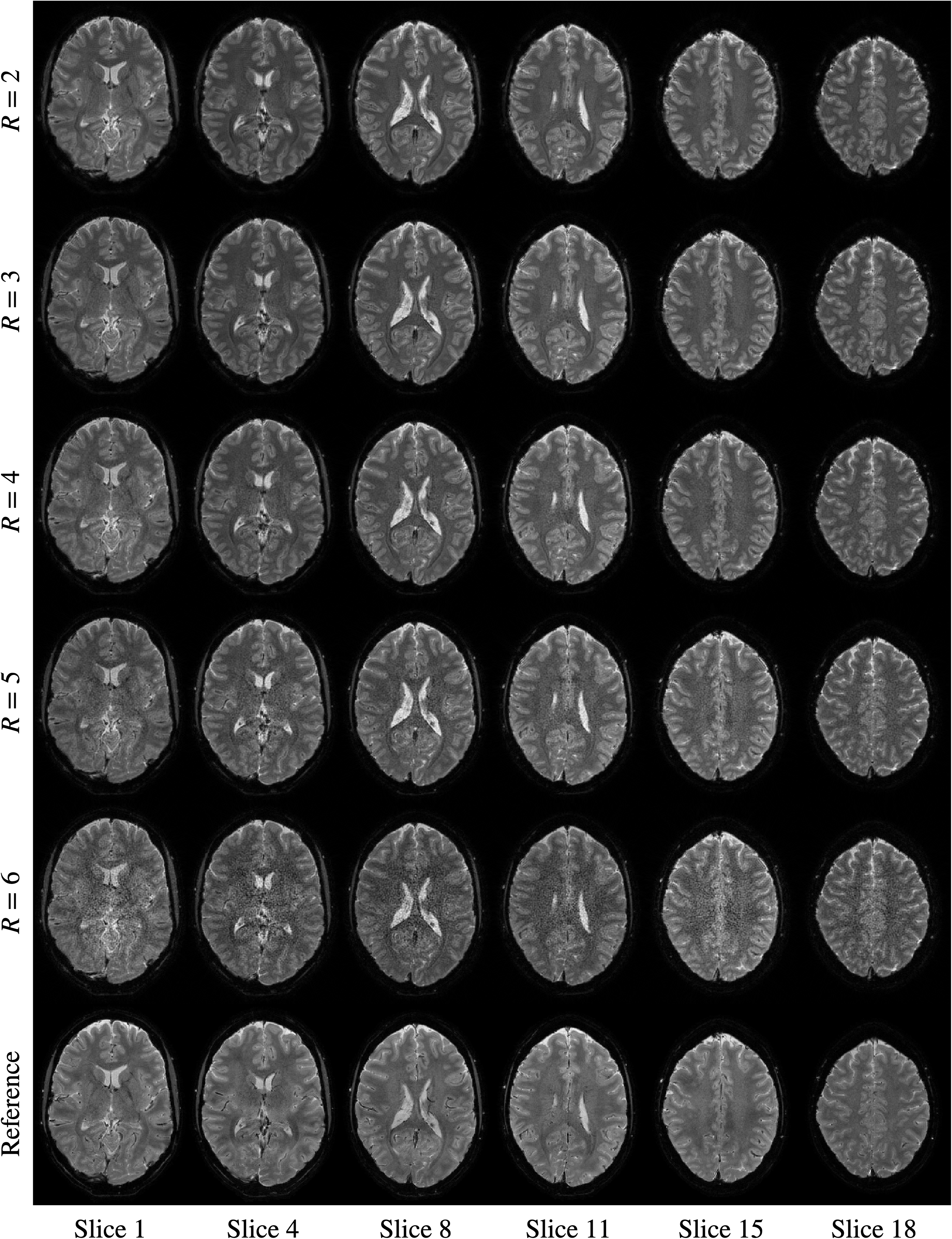}}
\caption{Single-shot 2D spiral images. FOV: $219$x$219$x$36mm$, Resolution: $1$x$1$x$2mm$, Slice gap: $0mm$, $R=2,3,4,5,6$, $T_E=30ms$, $T_{Aq}=71.5ms,47.7ms,35.8ms,28.7ms,24ms$, Number of CG iterations: $7,13,16,26,46$. Reference images from a spin-warp scan are shown at the bottom row for comparison.  \label{fig_spirals_2D}}
\end{figure*}
\newpage

\begin{figure*}
\centerline{\includegraphics[width=17cm]{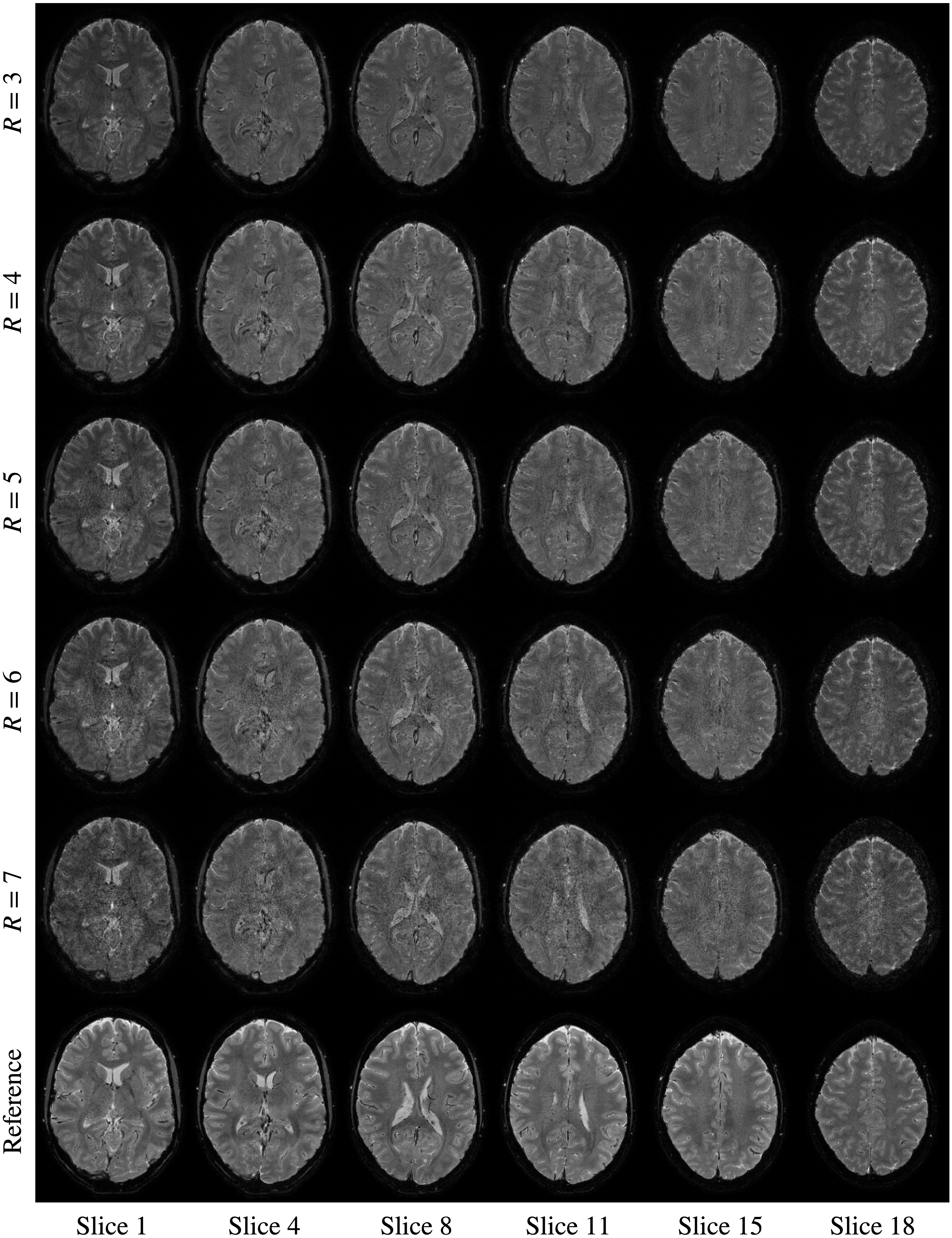}}
\caption{3D T-Hex spiral images. FOV: $219$x$219$x$36mm$, Resolution: $1$x$1$x$2mm$, $R=3,4,5,6,7$, $T_E=30ms$, $T_R=250ms$, $T_{Aq}=67.1ms,58.2ms,52.1ms,47.6ms,44.1ms$, Number of CG iterations: $47,37,46,50,56$. Reference images from a spin-warp scan are shown at the bottom row for comparison.  \label{fig_spirals_3D}}
\end{figure*}
\newpage

\begin{figure*}
\centerline{\includegraphics[width=17cm]{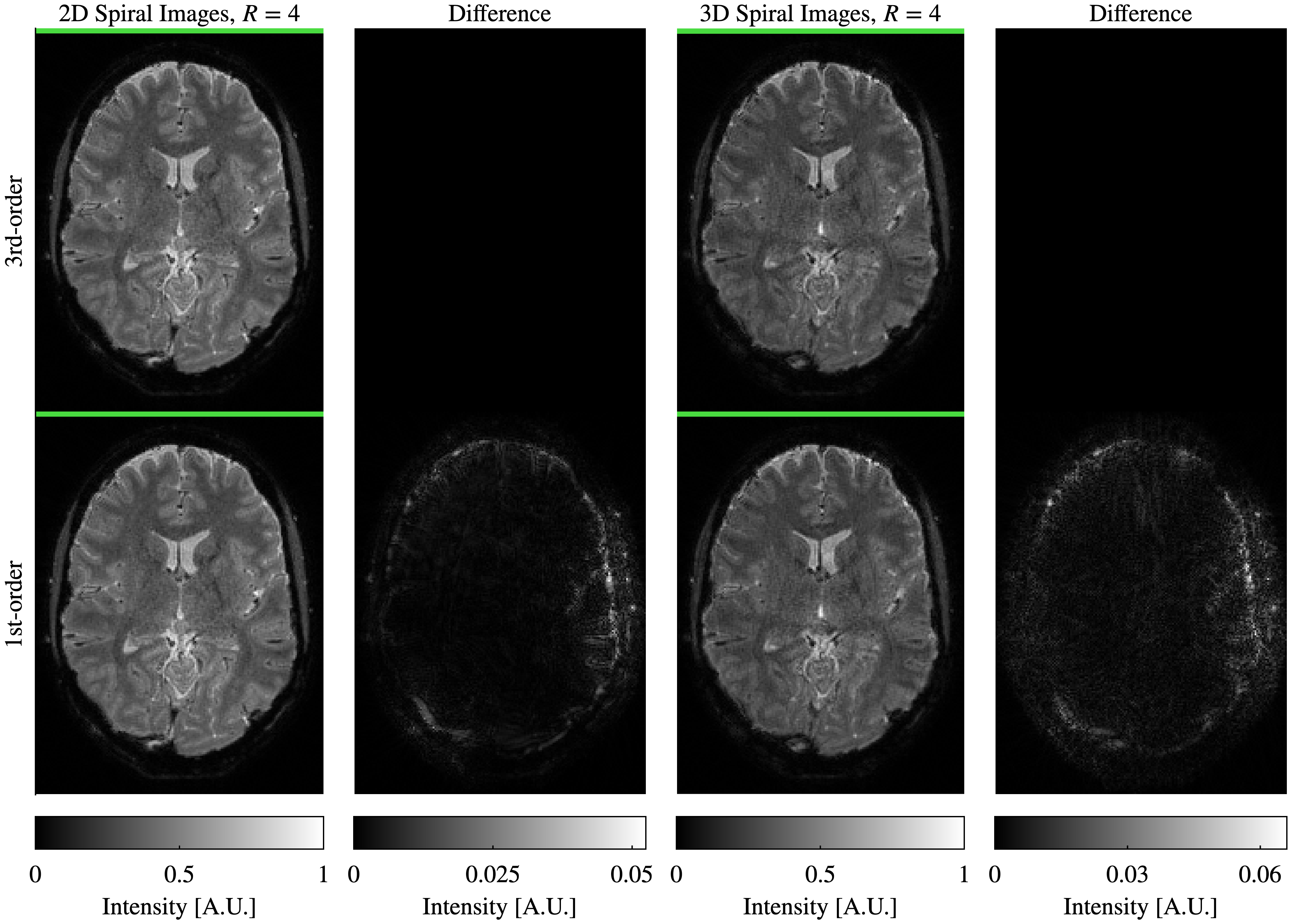}}
\caption{Influence of higher-order field terms on non-Fourier SENSE reconstruction. Top row: Reconstruction results when using field terms up to 3rd-order for the 2D and 3D spiral datasets. Bottom row: Equivalent results when using field terms up to 1st-order. \label{fig_higher_order_examples}}
\end{figure*}
\newpage

\begin{figure*}
\centerline{\includegraphics[width=17cm]{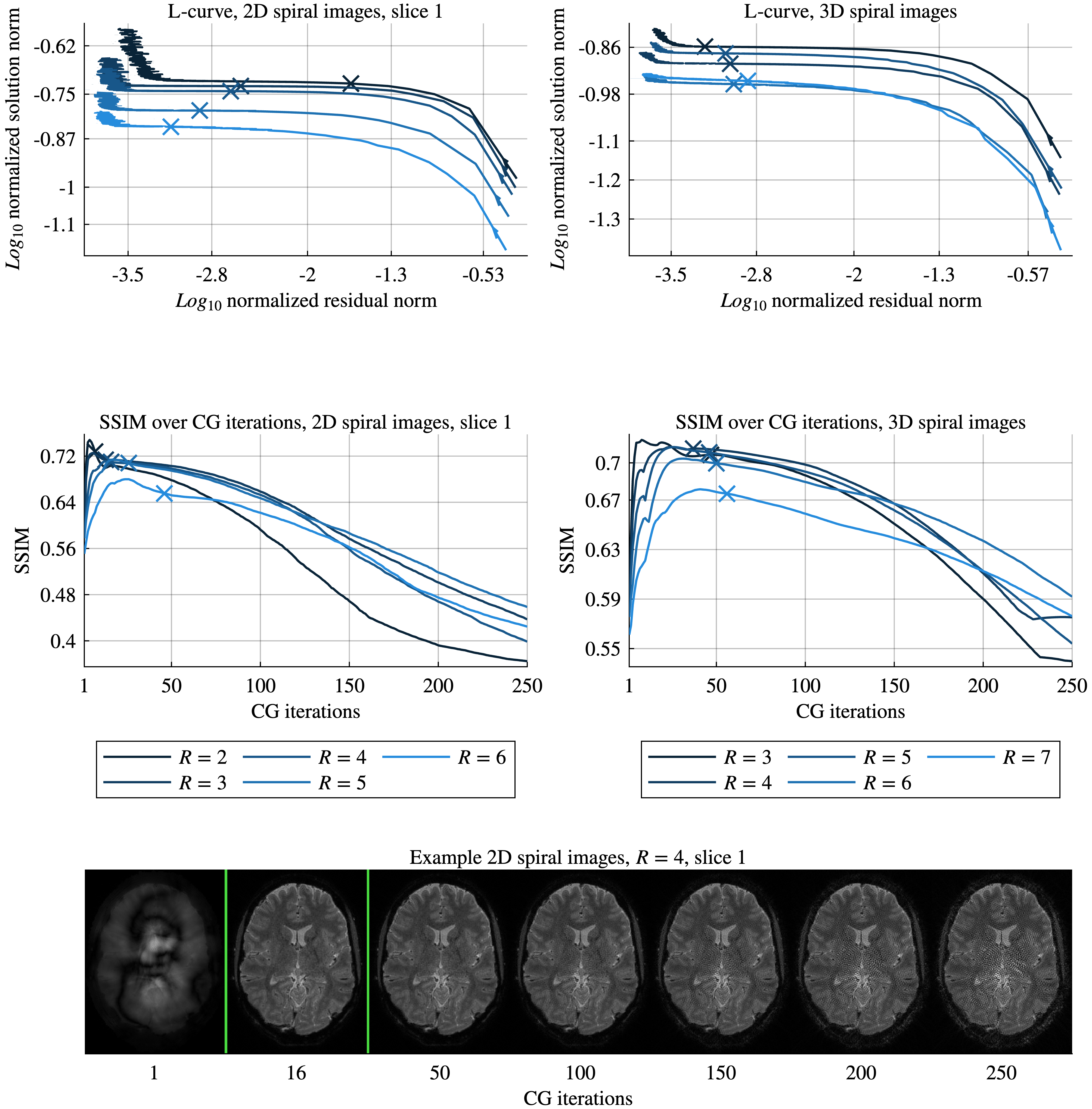}}
\caption{Top row: L-curves ($250$ CG iterations) for the reconstruction of $1$ slice of the 2D spiral dataset and the reconstruction of the 3D spiral dataset. Arrows indicate in which direction the L-curve is traversed over the CG iterations. Crosses indicate when visual inspection suggest good image quality. Middle row: Evolution of SSIM, for both datasets, over $250$ CG iterations. The reference image shown on the bottom of \myref{Figures }{fig_spirals_2D} and \myref{}{fig_spirals_3D} was used for comparison. Bottom row: Evolution of an example image over up to 250 iterations. \label{fig_cg_convergence}}
\end{figure*}
\newpage

\begin{figure*}
\centerline{\includegraphics[width=17cm]{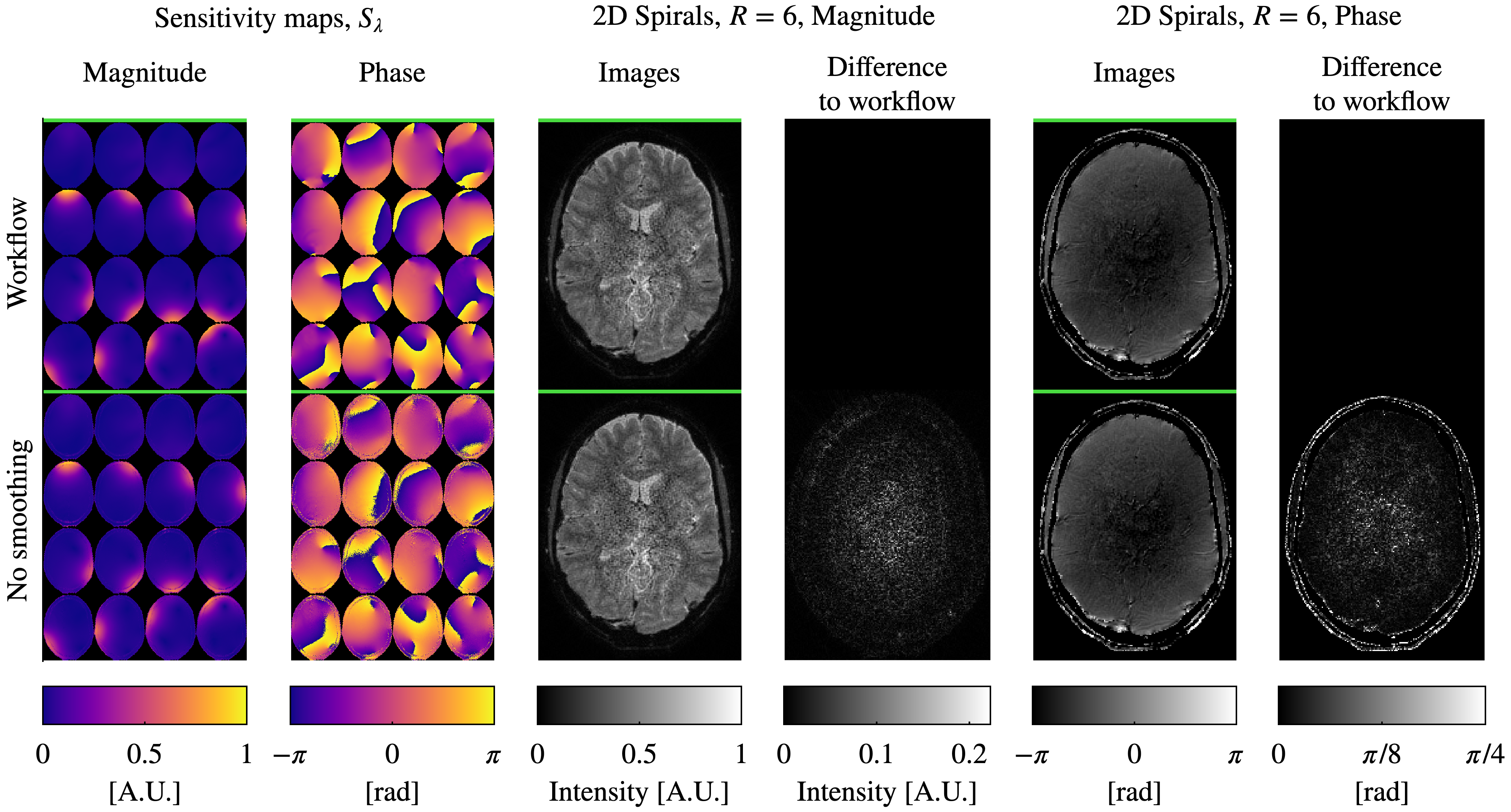}}
\caption{Influence of sensitivity map ($S_\lambda$) -processing on non-Fourier SENSE reconstruction. Top row: Sensitivity maps of all $16$ receiver coils and reconstruction examples after using the suggested smoothing and extrapolation algorithm. Bottom row: Equivalent examples computed without applying the suggested smoothing and extrapolation algorithm. Magnitude and phase of the reconstructed images are shown and compared individually.  \label{fig_sense_maps_examples}}
\end{figure*}
\newpage

\begin{figure*}
\centerline{\includegraphics[width=17cm]{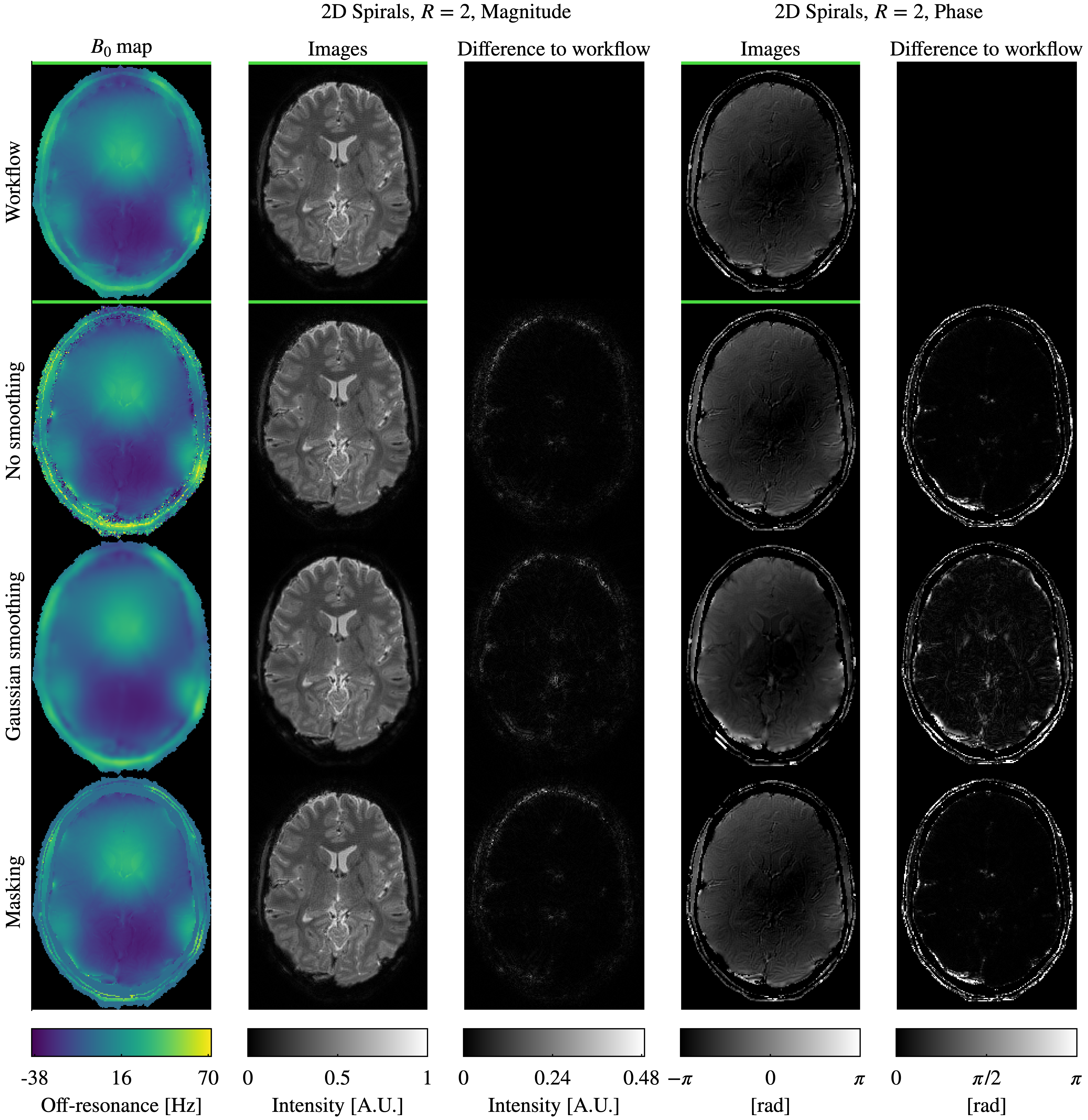}}
\caption{Influence of $B_0$ map-processing on non-Fourier SENSE reconstruction. Top row: $B_0$ maps and reconstruction examples after using the suggested smoothing and extrapolation algorithm. Second row: Equivalent examples computed without any processing of the $B_0$ map. Third row: Equivalent examples computed after smoothing the $B_0$ map with a Gaussian filter. Bottom row: Equivalent examples computed after removing noisy parts of the $B_0$ map by multiplying it with the trusted mask ($M_T$). Magnitude and phase of the reconstructed images are shown and compared individually.  \label{fig_b0_maps_examples}}
\end{figure*}
\newpage

\begin{figure*}
\centerline{\includegraphics[width=17cm]{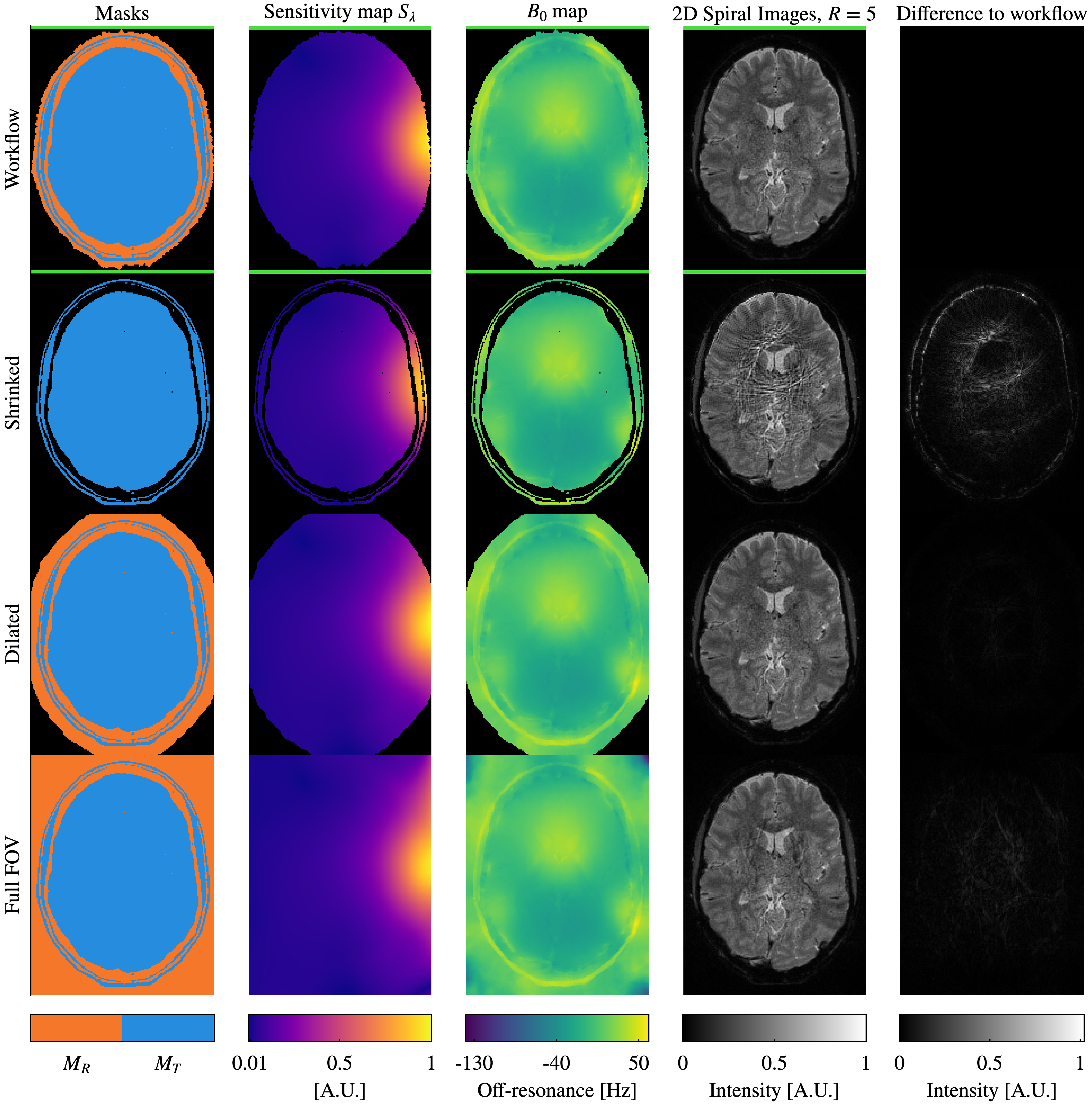}}
\caption{Influence of the reconstruction mask ($M_R$) on non-Fourier SENSE reconstruction. For the examples shown, $M_R$, computed as suggested, was shrinked to coincide with the trusted mask ($M_T$), dilated, and expanded to cover the full field of view (FOV). The corresponding masks are displayed in the left column. The sensitivity maps ($S_\lambda$) and $B_0$ maps were computed accordingly and are presented in the second and third columns, respectively. Reconstruction results for a 2D spiral acquisition ($R=5$) are shown in the fourth column (magnitude images only). Absolute difference images with respect to the suggested case are shown in the rightmost column.  \label{fig_maksing_examples}}
\end{figure*}
\newpage

\begin{figure*}
\centerline{\includegraphics[width=17cm]{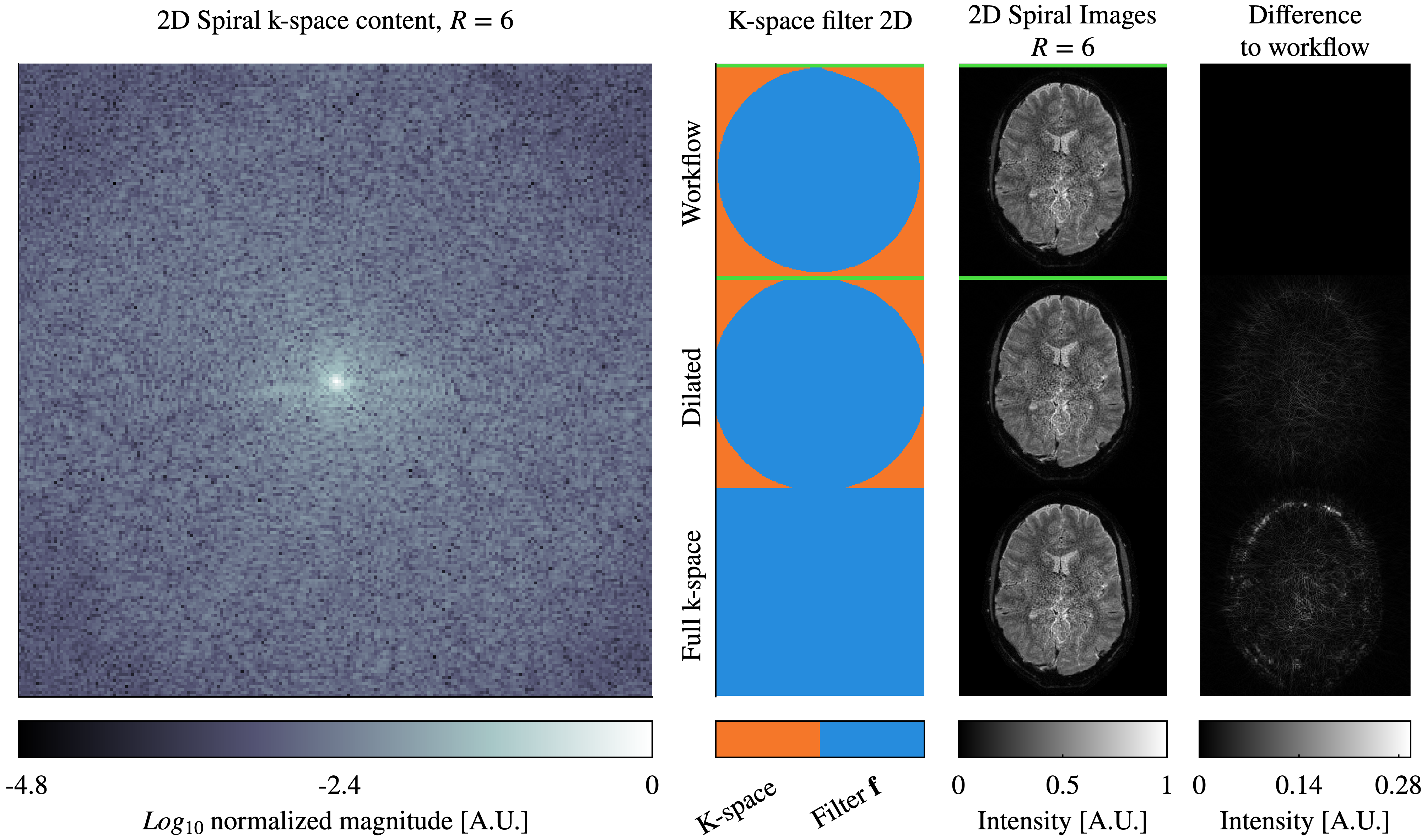}}
\caption{Influence of the k-space filter ($\mathbf{f}$) on non-Fourier SENSE reconstruction. Left side: Unfiltered k-space content of the reconstructed image. Right side, top row: K-space filter and reconstruction examples after using the suggested computation method. Second row: Equivalent examples computed after dilating the suggested k-space filter. Third row: Equivalent examples computed when not applying the kspace filter.\label{fig_kspace_filter_examples}}
\end{figure*}
\newpage

\FloatBarrier
\begin{table*}
\centerline{\includegraphics[width=17cm]{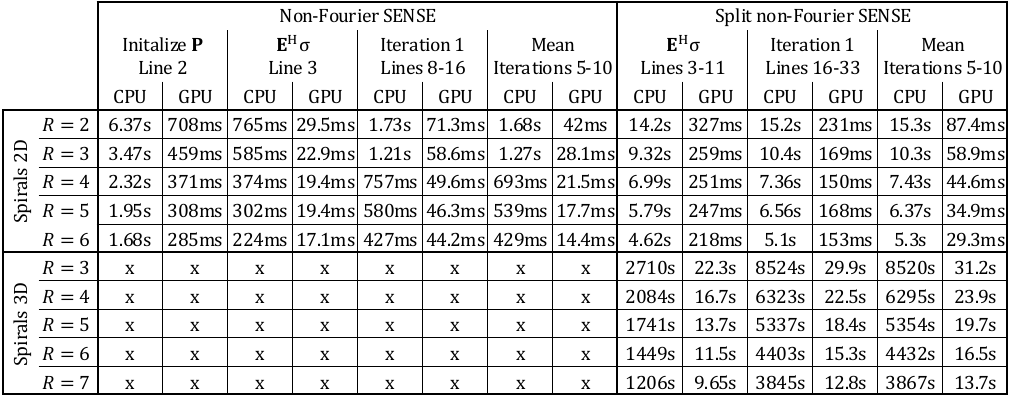}}
\caption{Runtime analysis of the proposed non-Fourier SENSE implementations. For the reconstruction of one slice of the 2D spiral dataset, measured runtimes are reported for both the non-Fourier SENSE and the split non-Fourier SENSE implementations. For the 3D spiral dataset, runtimes are provided for the split non-Fourier SENSE implementation. All reconstructions were performed on a 64-core AMD EPYC 7763 2.45 GHz CPU or a NVIDIA RTX 4090 24 GB GPU on a Windows 11 computer running Matlab 2023b. Line-numbers refer to the implementations as shown in \myref{Figure }{fig_recon_algorithms}. 
\label{table_runtimes}}
\end{table*}

\end{document}